\documentclass{elsart}

\usepackage{natbib}		%for referencing
\usepackage{amssymb}		%for symbols

  \usepackage{graphicx}
\def\url#1{{\ttfamily\def\/{/\discretionary{}{}{}}#1}}
\def\bibcode#1{(\texttt{#1})}

\newcommand{\SGZ}{{\rm SGZ}}

\newcommand{\Om}{\Omega_{M,0}}
\newcommand{\Ol}{\Omega_{\Lambda,0}}

\newcommand{\del}{\delta}

\newcommand{\hinv}{{h_0^{-1}}}
\newcommand{\mpc}{{\rm\,Mpc}}
\newcommand{\kpc}{{\rm\,kpc}}
\newcommand{\himpc}{\hinv{\rm\,Mpc}}
\newcommand{\hikpc}{\hinv{\rm\,kpc}}
\newcommand{\kms}{{\rm\,km\ s^{-1}}}
\newcommand{\kmsmpc}{{\rm\ km\ s^{-1}\ Mpc^{-1}}}

\newcommand{\Msun}{M_{\odot}}

\def\Fig#1{Figure~\ref{#1}}

\begin{document}

\begin{frontmatter}

\title{Future Evolution of Nearby Large-Scale Structure in a Universe
Dominated by a Cosmological Constant}

\author{Kentaro Nagamine\thanksref{ken}},
\author{Abraham Loeb\thanksref{avi}}
\address{Harvard-Smithsonian Center for Astrophysics, 
60 Garden Street, MS 51, Cambridge, MA 02138}

% use the thanksref command within \title, \author or \address for footnotes:
% \title{\thanksref{label1}}
% \thanks[label1]{}
% \author{\thanksref{label2}}
% \thanks[label2]{}
% \address{\thanksref{label3}}
% \thanks[label3]{}
% including your email address
% \address{\thanksref{email}}

\thanks[ken]{E-mail: knagamin@cfa.harvard.edu}
\thanks[avi]{E-mail: aloeb@cfa.harvard.edu}

%%%%%%%%%%%%%%%%%%%%%%%%%%%%%%%%%%%%%%%%%%%%%%%%%%%%%%%%%%%%%%%%%%%%%%

\begin{abstract}
We simulate the future evolution of the observed inhomogeneities in
the local universe assuming that the global expansion rate is
dominated by a cosmological constant.  We find that within two Hubble
times ($\sim 30$ billion years) from the present epoch, large-scale
structures will freeze in comoving coordinates and the mass
distribution of bound objects will stop evolving.  The Local Group
will get somewhat closer to the Virgo cluster in comoving coordinates,
but will be pulled away from the Virgo in physical coordinates due to
the accelerated expansion of the Universe.  In the distant future
there will only be one massive galaxy within our event horizon, namely
the merger product of the Andromeda and the Milky Way galaxies.  All
galaxies that are not gravitationally bound to the Local Group will
recede away from us and eventually exit from our event horizon.  More
generally, we identify the critical interior overdensity above which
a shell of matter around an object will remain bound to it at late
times.
\end{abstract}

\end{frontmatter}

%%%%%%%%%%%%%%%%%%%%%%%%%%%%%%%%%%%%%%%%%%%%%%%%%%%%%%%%%%%%%%%%%%%%%%
\section{Introduction}
\label{section:intro}

Recent data on the temperature anisotropies of the cosmic microwave
background \citep[e.g.][]{Hanany00, Bernardis00}, the luminosity distance
to Type Ia supernovae \citep{Perlmutter98, Riess98, Garnavich98}, and the
large-scale distribution of galaxies \citep{Peacock01, Verde02} favors a
flat universe with present density parameters of $\Om=0.30\pm 0.15$ in matter 
and $\Ol=1-\Om$ in a cosmological constant. The energy density of the vacuum
(the so-called cosmological constant or ``dark energy'') appears to be
currently dominating the expansion rate of the Universe.

Given these specific values for the cosmological density parameters,
it has now become possible to predict quantitatively the future
evolution of the visible Universe.  The existence of a cosmological
constant has profound consequences in this context \citep[see,
e.g.][]{Staro00, Gud02, Loeb02}; in particular: (i) when the Universe
will age by a factor of a few, the event horizon will stall at a fixed
proper distance of $3.6h_0^{-1}$ Gpc around us (where $h_0$ is the
present Hubble constant in units of $100\kmsmpc$); (ii) all sources
with present redshifts larger than 1.8 have already crossed our event
horizon and are therefore out of causal contact today; and (iii) even
if we continue to monitor the sources with current redshifts of
$z=5-10$ into the infinite future, we will only be able to see these
sources acquire intrinsic ages of $4-6$ billion years in their rest
frame.

In this paper we use N-body simulations to calculate the future
gravitational growth of the observed large-scale structure in the
local universe. For simplicity, we assume that the vacuum energy
density is constant in time (as for a classical cosmological constant)
and adopt the values $\Om=0.3$, $\Ol=0.7$ and $h_0=0.7$.  We simulate
the evolution of density inhomogeneities within a sphere of radius
$\sim 100\himpc$ around the Milky Way galaxy.  Our goal is to
determine the time when the accelerated expansion of the Universe will
freeze the large-scale structures in distant future, as well as to
find the minimum overdensity of matter interior to a shell surrounding
an object today that will allow the shell to remain bound to the
object at late times despite the repulsive gravitational force of the
vacuum.  Another issue of particular interest is the question whether
the Local Group of galaxies is bound to the nearby Virgo cluster.  It
has been well-established that the Local Group has a peculiar infall
velocity towards the direction of the Virgo cluster, the so-called
`Virgocentric Infall' \citep[e.g.][]{Aaronson82, Davis83,
Lyndenbell88}, but it has not been investigated yet by a direct
numerical simulation whether the Local Group will eventually fall into
the Virgo cluster in the future given the observed galaxy distribution
in the local universe.  The possibility remains that the Virgocentric
Infall will freeze in comoving coordinates due to the exponential
expansion of the universe, and the Local Group will not ultimately
merge with the Virgo cluster.  We investigate this issue for the first
time by running an N-body simulation, starting from initial conditions
at z=0 that match the observed galaxy distribution in the nearby universe 
and ending it in a distant future.

The outline of this paper is as follows. In \S~\ref{section:simulation} we
describe the characteristics of our N-body simulations for the evolution of
the global structure of the local universe. The results from the
simulations are discussed in \S~\ref{section:evolution}. We then derive in
\S~\ref{section:bound} the present-day overdensity threshold for a
spherical region to collapse in the future, and apply this condition to the
Local Group in \S~\ref{section:LG}. Finally, we summarize our main
conclusions in \S~\ref{section:conclusion}.

%%%%%%%%%%%%%%%%%%%%%%%%%%%%%%%%%%%%%%%%%%%%%%%%%%%%%%%%%%%%%%%%%%%%%%

\section{Simulations}
\label{section:simulation}

The simulations were carried out with the parallel tree N-body/SPH
code GADGET\footnote{\url{http://www.MPA-Garching.MPG.DE/gadget/}}
\citep{Springel01}.  The SPH component of the code was turned off as
we are only interested in the dark matter particles which dominate the
fluctuating mass density.  For the initial conditions, we used the
simulation output of
\citet{Mathis02}\footnote{\url{http://www.MPA-Garching.MPG.DE/NumCos/CR/}}
at $z=0$.  This simulation was designed so that at $z=0$ the simulated
mass density field would best match the observed galaxy overdensity
distribution in the {\it IRAS} 1.2 Jy survey of \citet{Fisher94,
Fisher95}.  \citet{Mathis02} constructed mock catalogues of the Mark
III Catalogue of Peculiar Velocities \citep{Willick95, Willick96,
Willick97} using their simulation result, and have shown that the
radial peculiar velocities in mock and real catalogues agree very
well.  The original data set of \citet{Mathis02} contains 50.7 million
dark matter particles of mass $3.6\times 10^9\hinv\Msun$ within the
comoving radius of $80\himpc$ around the supergalactic center (inner
high-resolution region), and 20.5 million particles within the
comoving radius of $\sim 200\himpc$ (outer low-resolution region). The
simulation of this spherical region was run with a vacuum boundary
condition.

Because we are only interested in the evolution of the structure with
mass-scale larger than that of the Local Group, we reduce the original
data set by a factor of 100 through random sampling of particles.
This reduction of data allows us to run the simulation with a
relatively low computational cost while preserving the mass and
spatial resolution we need to make predictions for the questions we
are interested in, given the observed galaxy distribution in the nearby
universe.  The resulting mass of each dark matter particle in the high
resolution region is $3.6\times 10^{11}\hinv\Msun$ (comparable to the
mass of an $L_\star$ galaxy), and the number of high resolution
particles is half a million.  We have set the gravitational softening
length to the physical scale of $100\hinv\kpc$, since structures below
this scale are not of interest.  This mass and spatial resolution is
sufficient for following the overall structure of the local universe,
but not sufficient for resolving the inner details of galaxy groups
such as the Local Group, of which the Milky Way galaxy is the dominant
member.

Using the reduced data set, we have simulated the evolution of the
local universe from the present time (corresponding to a scale factor value
of $a=1$ at $z=0$) to the 6 Hubble times into the future ($a=166$; $\sim
84$ billion years from today).  Because the present Hubble time $t_H\equiv
1/H_0=14$ Gyr (for $h_0=0.7$) is very close to the present age of 
the Universe $t_0=13.5$ Gyr in the adopted cosmology, the epoch of $a=166$ 
roughly corresponds to $t=t_0 + 6t_H \approx 7t_H$ from the Big Bang ($t=0$).  
Note that we denote all present-day values with a subscript zero.

In addition to the above constrained realization simulation of the local
universe, we have also performed a simulation with a periodic box of ${\rm
L}=100\himpc$ containing ${\rm N}=64^3$ dark matter particles, in order to
make sure that the statistical results presented in this paper are not
affected by cosmic variance or by the choice of the vacuum boundary
condition for the constrained realization run.  Statistically, we find that
the results from the two runs agree very well with one another, and so we
only show the results from the constrained realization run in
\S~\ref{section:evolution}.

The simulations were performed on the local Beowulf PC cluster located at
Harvard-Smithsonian Center for Astrophysics.  It is interesting to note
that the numerical code speeds up its calculation at late cosmic times.
The accelerated expansion of the Universe freezes the large-scale structures 
in comoving coordinates, and so at late times the variable time step of the
code increases and the tree update for the force computation becomes less
frequent.  
%%Using 16 Pentium III CPUs, each run was completed within a day.

%%%%%%%%%%%%%%%%%%%%%%%%%%%%%%%%%%%%%%%%%%%%%%%%%%%%%%%%%%%%%%%%%

\section{Future Evolution of Large-Scale Structure in the Local Universe}
\label{section:evolution}

\Fig{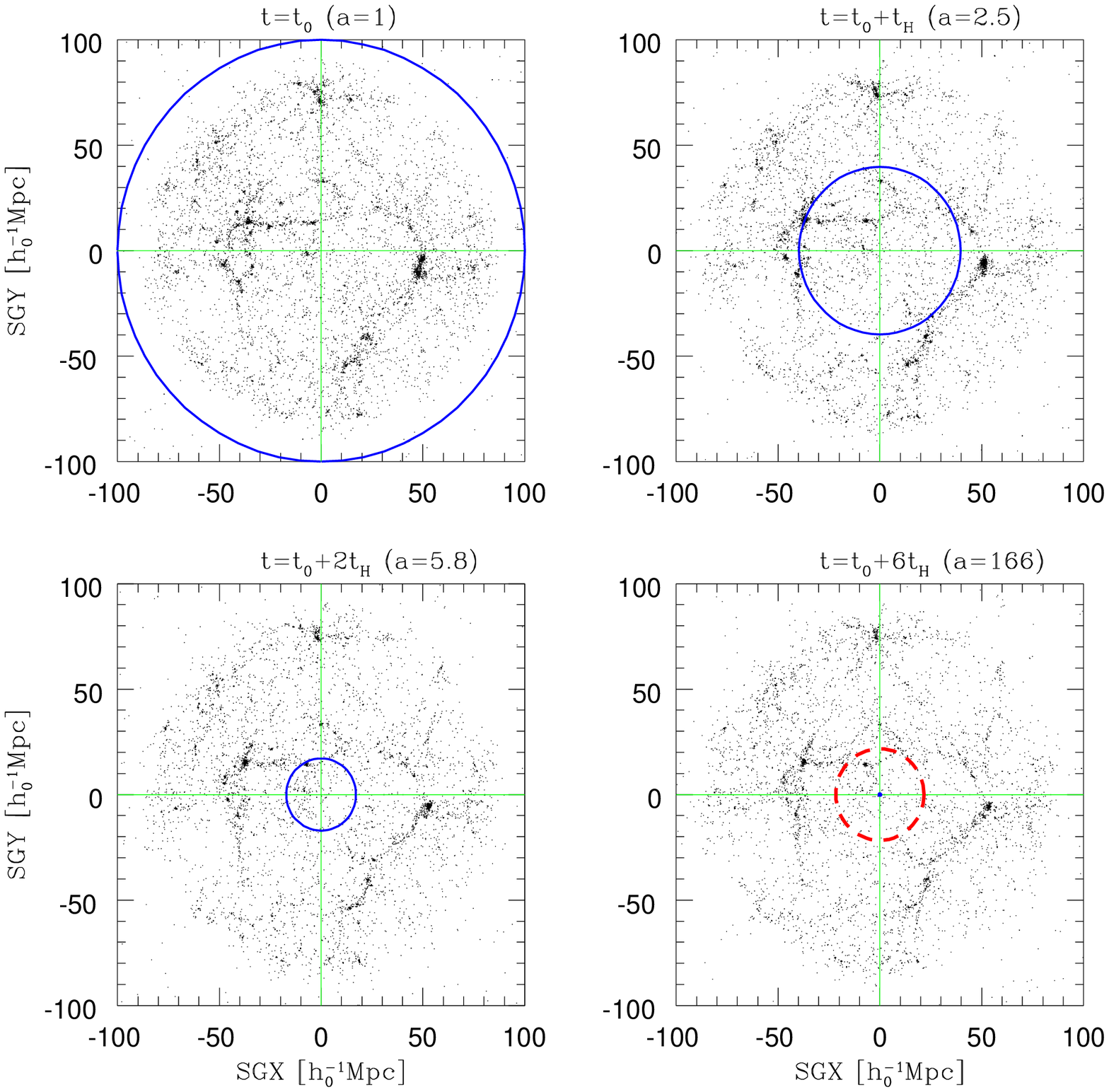} shows the distribution of particles in a slab covering a width
(centered at zero) along the supergalactic Z axis of $-15<\SGZ<15\himpc$,
and projected onto the supergalactic XY plane in comoving coordinates. From
top left to bottom right the four panels show snapshots at times $t=t_0$,
$t_0+t_H$, $t_0+2t_H$, and $t_0+6t_H$, corresponding to $a=1.0$, 2.5, 5.8,
and 166, respectively.  The solid circle in each panel denotes the physical
radius of $100\himpc$ at each epoch. At $t=t_0+6t_H$, the Universe 
has expanded so much that this circle is no longer visible. 
Three major density peaks can be
easily identified in all panels: the Coma cluster at the top, the Great
Attractor (Centaurus) to the left of the center, and the Perseus-Pisces to
the right of the center. The Virgo cluster is also apparent just above the
center.  Close examination of the structures reveals that most of the
evolution takes place between $t_0$ and $t_0+2t_H$, and little evolution is
seen after $t=t_0+2t_H$.  This follows from the fact that after a few
Hubble times, the exponential growth of the scale factor, $a(t)\propto
\exp(\sqrt{\Ol}t/t_H)$, damps any peculiar velocity which is
gravitationally induced by mild density inhomogeneities. The bottom right
panel shows our event horizon by the thick dashed circle around the 
supergalactic center, which is located at a physical radius of
$3.6h_0^{-1}$ Gpc.  At $t=t_0+6t_H$, all clusters except Virgo are already
outside our event horizon. The Virgo cluster will also exit from the event
horizon at $t\approx t_0+6.3t_H$ ($a=210$; $\sim 88$ billion years from today) 
and its image will fade quickly while
remaining frozen on the sky at the time of its exit.  Although our
calculation does not take general relativistic effects into account, we
expect our near-horizon results to be valid, since structures on this scale
should at any event be frozen.

\begin{figure}
\begin{center}
\includegraphics*[scale=1, width=40pc, height=40pc]{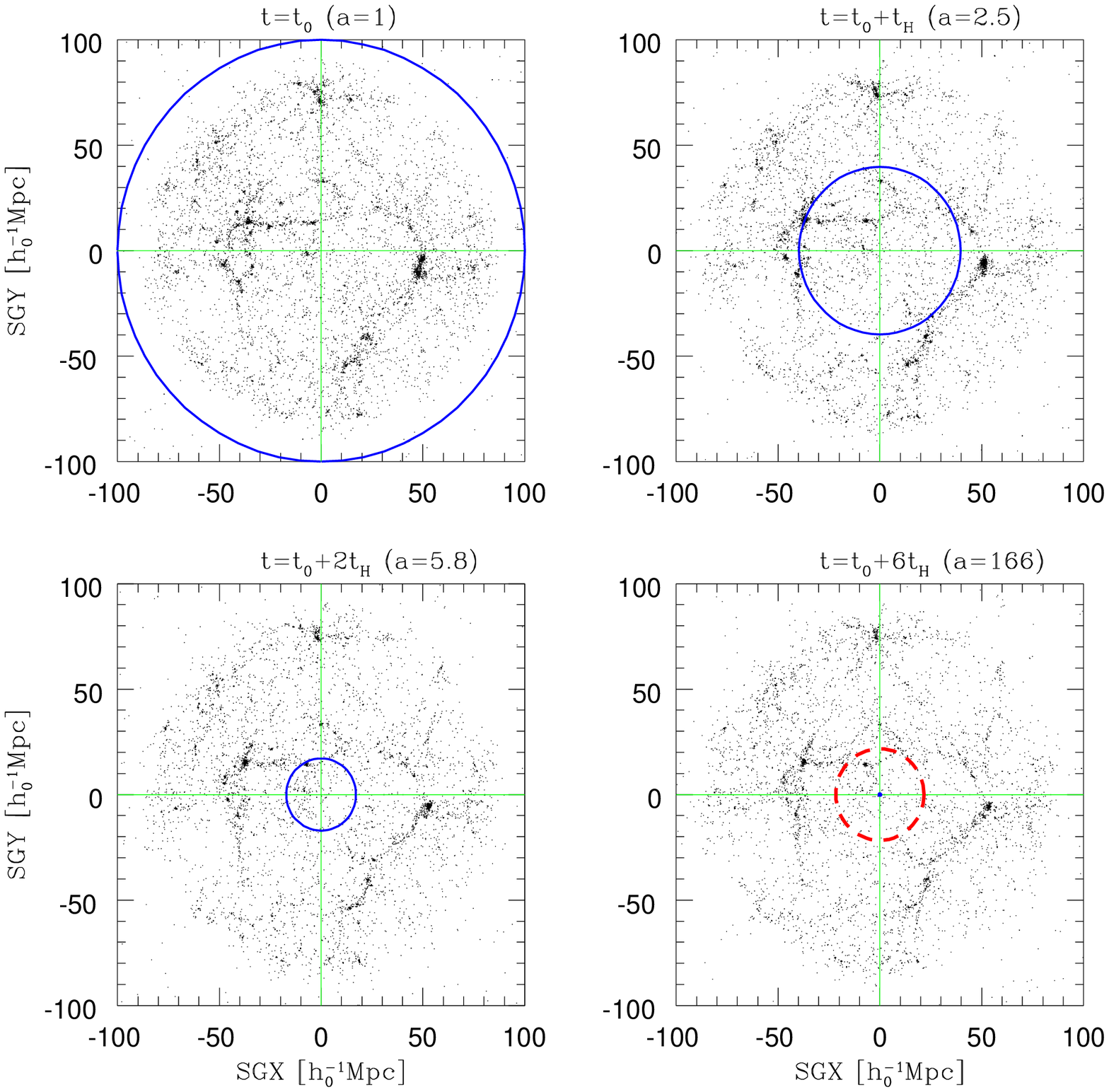}
\end{center}
\caption{Future evolution of the local universe in comoving coordinates.
Shown are the particles in a slab of thickness $-15<\SGZ<15\himpc$ projected
onto the supergalactic XY plane at the times $t=t_0$, $t_0+t_H$,
$t_0+2t_H$, and $t_0+6t_H$ (corresponding to $a=1.0$, 2.5, 5.8, and 166)
from top left to bottom right. The thick solid circle in each panel
indicates the physical radius of $100\himpc$ around the supergalactic
center.  In the bottom right panel, the Universe has expanded so much that
this circle is no longer visible. Instead, we show the event horizon at a
physical radius of $3.6h_0^{-1}$ Gpc as the thick dashed circle.
[The number of particles in this figure is further reduced by a factor
of 20 from the original version for the astro-ph submission. See 
http://cfa-www.harvard.edu/~$\tilde{}$ knagamine/LocalGroup  ~for high 
resolution figures.]
\label{f1.eps}}
\end{figure}

The lack of late evolution is even more apparent in the mass distribution
of bound objects.  \Fig{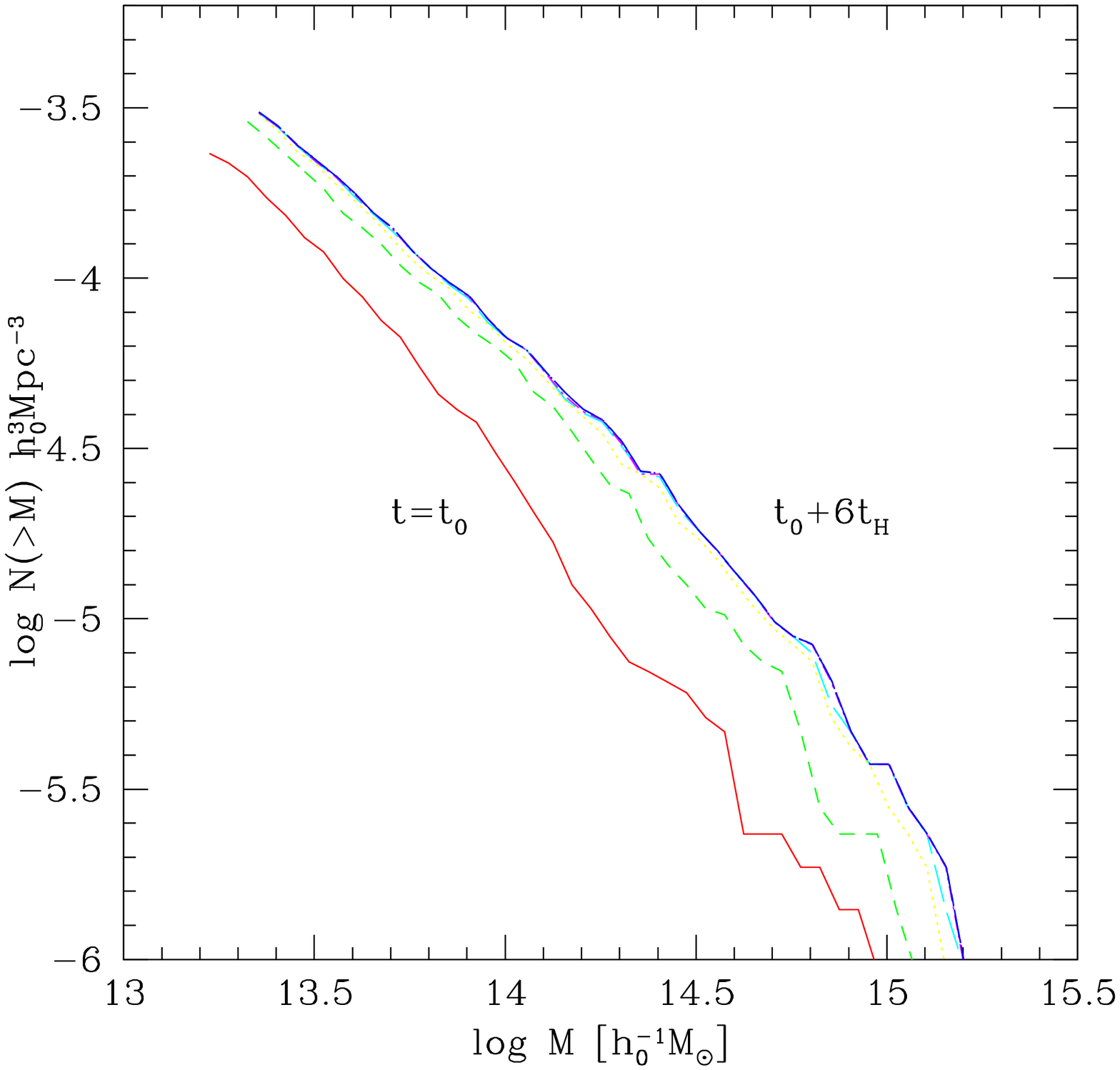} shows the cumulative number of dark matter
halos with mass above $M$ per comoving volume, N($>$$M$).  The dark matter
halos are identified through the HOP grouping algorithm
\citep{Eisenstein98} with the outer overdensity threshold of $\del_{\rm
out}=80$ and the peak overdensity threshold of $\del_{\rm peak}=240$ in
comoving coordinates. From left to right, the lines refer to times
$t=t_0$ ({\it solid}), $t_0+t_H$ ({\it short-dashed}), $t_0+2t_H$ ({\it
dotted}), $t_0+3t_H$ ({\it long-dashed}), $t_0+4t_H$ ({\it dot-short
dashed}), $t_0+5t_H$ ({\it dot-long dashed}), and $t_0+6t_H$ ({\it short
dash - long dash}), which correspond to scale factor values of $a=1.0$,
2.5, 5.8, 13.5, 31.2, 72.0, and 166, respectively.  The last four lines
are almost identical within sampling errors, showing that the mass function
freezes after two Hubble times from today and that mergers are very rare
subsequently.

\begin{figure}
\begin{center}
\includegraphics*[scale=1, width=33pc, height=33pc]{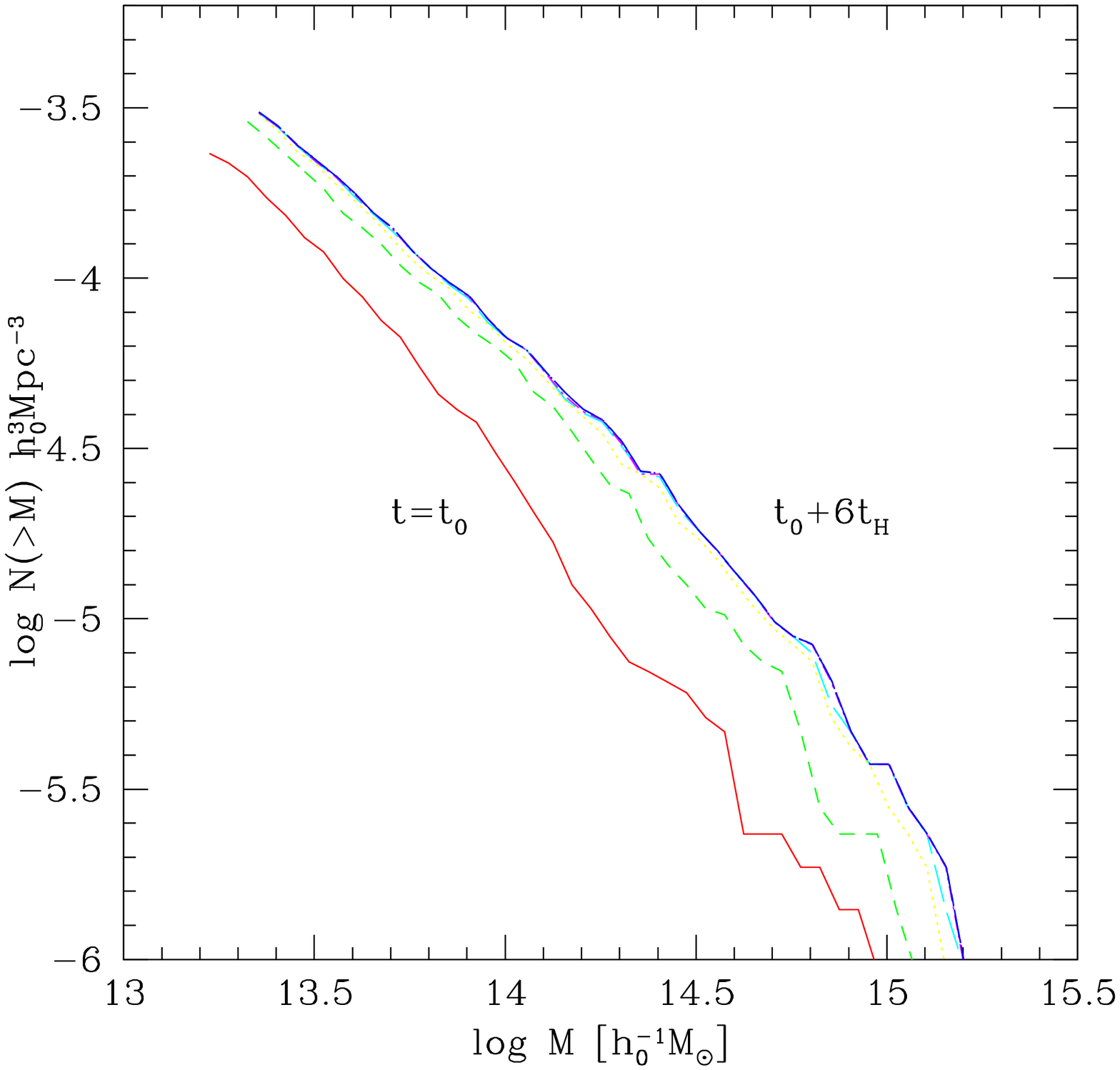}
\end{center}
\caption{Evolution of the cumulative number of dark matter halos with mass
above $M$ per comoving volume, N($>$$M$). From left to right, $t=t_0$ ({\it
solid}), $t_0+t_H$ ({\it short-dashed}), $t_0+2t_H$ ({\it dotted}),
$t_0+3t_H$ ({\it long-dashed}), $t_0+4t_H$ ({\it dot-short dashed}),
$t_0+5t_H$ ({\it dot-long dashed}), and $t_0+6t_H$ ({\it short dash - long
dash}), which correspond to $a=1.0$, 2.5, 5.8, 13.5, 31.2, 72.0, and 166,
respectively. The last four lines are almost indistinguishable, implying
that the mass function freezes after $\sim 2t_H$ from the present time and
the merger or accretion rates are very low subsequently.
\label{f2.eps}}
\end{figure}

%%%%%%%%%%%%%%%%%%%%%%%%%%%%%%%%%%%%%%%%%%%%%%%%%%%%%%%%%%%%%%%%%%%%%%

\section{Gravitationally Bound Objects}
\label{section:bound}

Next we consider the interior overdensity threshold above which a shell of
matter around a spherically symmetric object will remain bound to it at
late times. In the absence of a cosmological constant and with no shell
crossing, the threshold simply amounts to the kinetic$+$potential energy of
the shell being negative \citep{Gunn72}.  The existence of a cosmological
constant makes the interior gravitating mass of the vacuum grow larger as
the shell radius increases; consequently a shell with a negative energy
today may still be pulled away from the object by the cosmic acceleration
at late times. We would like to find the minimum mean interior overdensity
today, $\del_c$, above which the shell will remain bound to the object at
arbitrarily late times. Here $\delta= (\rho-\bar{\rho})/\bar{\rho}$, where
$\bar{\rho}=\Om (3H_0^2/8\pi G)$ is the mean density of matter in the
Universe.

The critical mean overdensity interior to the shell radius, $\delta_c$, was
derived by \citet{Lokas02} in their equation (27). Taking the initial time
as $z=0$, this equation yields $\del_c=17.6$ for the set of cosmological
parameters adopted in this paper.  We note that \citet{Lokas02} have
ignored the possibility that the mass shell may have a nonzero initial
peculiar velocity.  For the growing mode of a density perturbation, the
inward peculiar velocity will induce collapse at a lower initial
overdensity than that found by \citet{Lokas02}.

\begin{figure}
\begin{center}
\includegraphics*[scale=1, width=33pc, height=30pc]{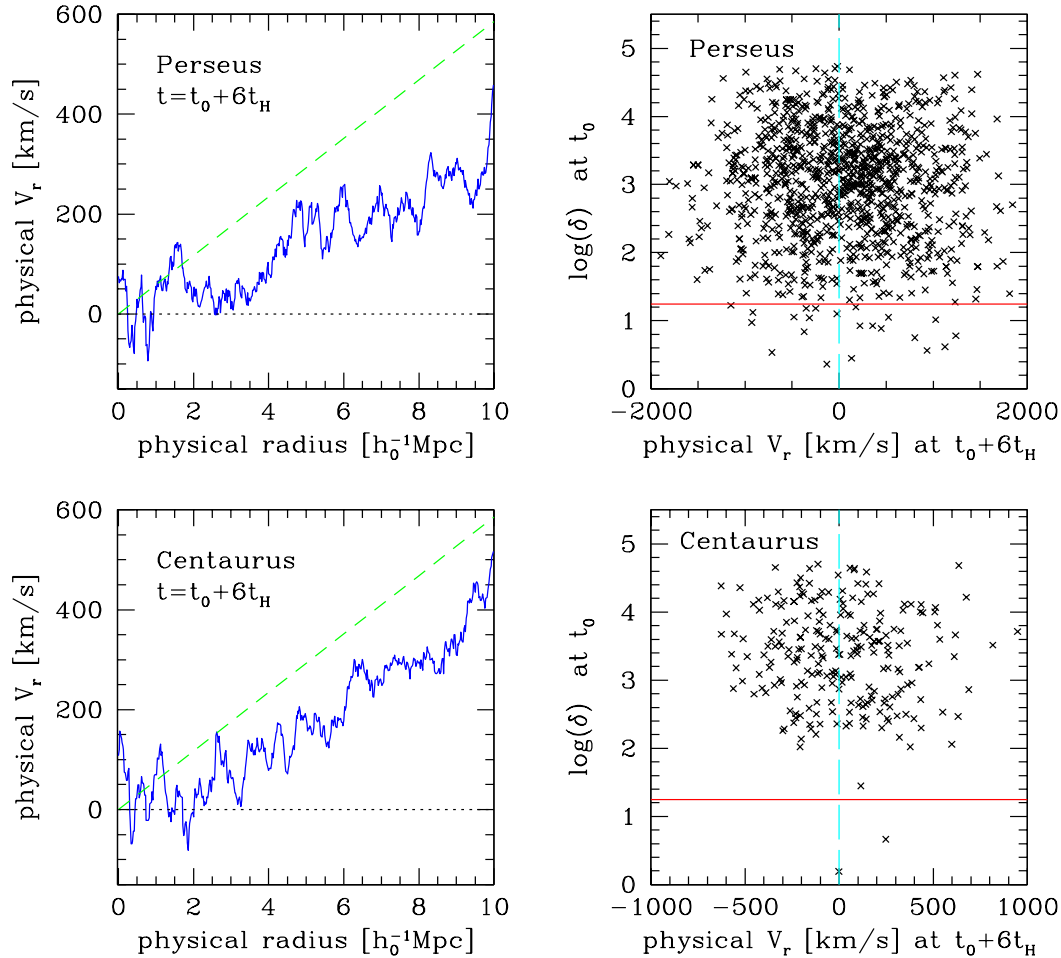}
\end{center}
\caption{{\it Left column}: mean physical radial velocity $v_r$ of
concentric shells around the maximum density peak is shown as a function of
physical radius at $t=t_0+6t_H$. The short-dashed line indicates the Hubble
law at $t=t_0+6t_H$.  The velocity fluctuations around zero inside the
radius of $2\himpc$ from the cluster center indicate that the associated
particles have virialized.  {\it Right column}: the present-day overdensity
of particles that are within the physical radius of $2\himpc$ around the
maximum density peak at $t=t_0+6t_H$ versus the physical radial velocity of
these particles at $t=t_0+6t_H$. Zero velocity is indicated by the vertical
long-dashed line, and the critical overdensity $\del_c=17.6$ is indicated
by the solid horizontal line.
\label{bound.eps}}
\end{figure}

We have attempted to test the validity of the critical overdensity
threshold $\del_c=17.6$ with our N-body simulation. \Fig{bound.eps}
shows the physical radial velocity of particles in and around the two
most massive objects in the simulation: Perseus and Centaurus
clusters.  For the HOP grouping parameter of $\del_{\rm out}=80$ (see
\S~\ref{section:evolution}), the total grouped masses of Perseus and
Centaurus are $1.6\times 10^{15}\hinv\Msun$ and $9.0\times
10^{14}\hinv\Msun$ at $t=t_0$, and $5.5\times 10^{15}\hinv\Msun$ and
$3.0\times 10^{15}\hinv\Msun$ at $t=t_0+6t_H$, respectively (see \S~4
of \citet{Mathis02} for a detailed comparison between the simulated
and observed cluster properties).  On the left column of
\Fig{bound.eps} we show the mean physical radial velocity of
concentric shells around the density maximum as a function of their
physical radius at $t=t_0+6t_H$. The short-dashed line indicates the
Hubble law at $t=t_0+6t_H$, and the horizontal dotted line indicates
zero velocity.  Inside the radius of $2\himpc$ from the cluster center
the velocity fluctuates around zero, implying that the particles have
virialized. At larger radii, the velocity field approaches the Hubble
flow (denoted by the dashed line).  Because the physical radius of
$10\himpc$ at $t=t_0+6t_H$ corresponds to comoving $60\hikpc$, dark
matter particles at that radius are still infalling towards the
density peak and so the physical radial velocity does not perfectly
match the Hubble flow.

On the right column of \Fig{bound.eps} we have selected out all particles
that are within a physical radius of $2\himpc$ around the density maximum
at $t=t_0+6t_H$; these particles can be regarded as bound and virialized
based on the left column plots.  We then trace back the overdensity that 
these particles had at $t=t_0$, and plot it against the radial particle
velocity at $t=t_0+6t_H$.  Zero velocity is indicated by the vertical
long-dashed line, and the critical overdensity $\del_c=17.6$ is indicated
by the solid horizontal line. For a small number of particles, the value of
$\del_c=17.6$ is an overestimate, probably due to them having inward
initial peculiar velocities.  On average, the analytic estimate for the
critical overdensity based on the spherical tophat collapse model appears
to provide a good approximation to the actual threshold.

%%%%%%%%%%%%%%%%%%%%%%%%%%%%%%%%%%%%%%%%%%%%%%%%%%%%%%%%%%%%%%%%%%%%%%

\section{Future Evolution of the Local Group}
\label{section:LG}

\subsection{Will the Local Group Fall into Virgo?}

\begin{figure}
\begin{center}
\includegraphics*[scale=1, width=35pc, height=18pc]{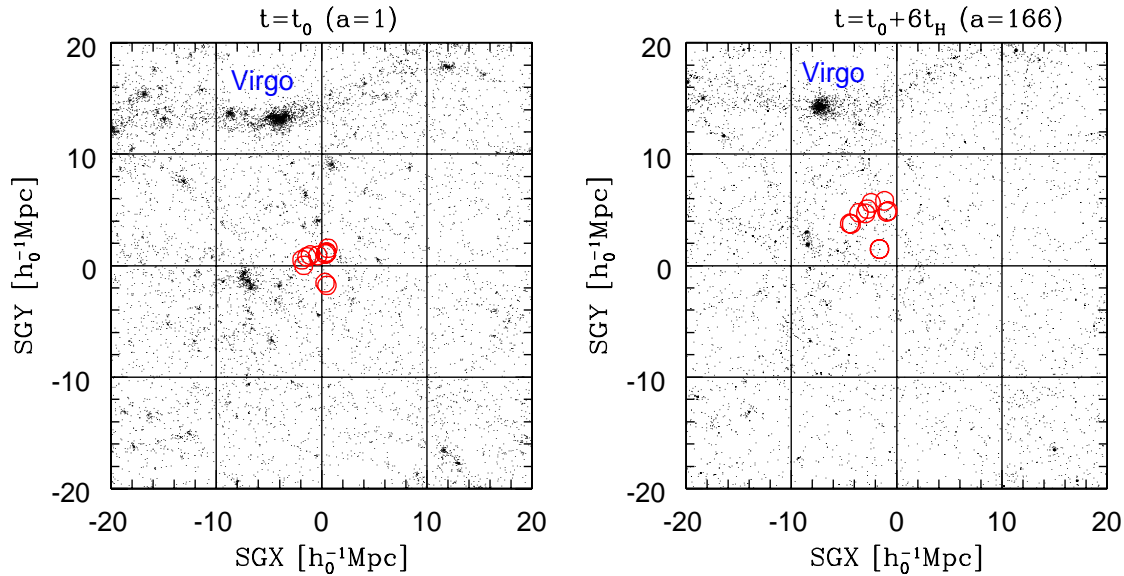}
\end{center}
\caption{Particle distribution in a slab of thickness $-15<\SGZ<15\himpc$
projected onto the supergalactic XY plane. The left panel shows the
projection today, and the right panel shows it at $t=t_0+6t_H$.  The Virgo
cluster appears near the top of the image.  Open circles mark particles
that are within a radius of $2\himpc$ from the supergalactic center at the
present time. They advance towards Virgo by a comoving distance of $\sim 5
\himpc$, but do not fall into it at late times. Most of this displacement
is completed by $t=t_0+2t_H$.
\label{virgo.eps}}
\end{figure}

Although our simulation does not resolve the inner structure of the
Local Group, we can get an idea about the future trajectory that its
center-of-mass will follow by tracing the particles which reside near
the supergalactic center at the present time. In \Fig{virgo.eps}, we
show the projection of a region with $-15<\SGZ<15\himpc$ onto the
supergalactic XY plane in a similar fashion to \Fig{f1.eps}.  The
density peak near the top of the panel is the Virgo cluster, which is
separated by a comoving distance of $14\himpc$ from the supergalactic
center.  The observational estimates of the distance to the Virgo
cluster from us by the methods of surface-brightness fluctuations,
planetary nebula luminosity function, Tully-Fisher relation, and
Cepheid variable stars, all agree with $16\pm 2\mpc$
\citep[e.g.][]{Jacoby92, Graham99}, in reasonable agreement with
the simulated distance within $2-\sigma$.

We identified all particles enclosed within a sphere of radius
$2\himpc$ around the supergalactic center at the present time as shown
in the left panel of \Fig{virgo.eps}). We have found 11 such particles
(shown as the open circles), corresponding to a total mass of
$3.9\times 10^{12}\hinv\Msun$.  For comparison, the Local Group has a
zero-velocity surface (separating it from the Hubble expansion) of
$1.18\pm 0.15\mpc$ and a total mass of $M_{\rm LG}=(2.3\pm 0.6)\times
10^{12}\Msun$ \citep{Courteau99, Bergh99} based on the radial velocity
dispersion of its members.  Initially, the sampled 11 particles have a
mean peculiar velocity of $670\kms$ at $t=t_0$ towards the direction
of $(-250, 502, -368)\kms$ with respect to the cosmic microwave
background (CMB) frame.  For comparison, the observed motion of the
Local Group relative to the CMB frame is $627\pm 22\kms$ in the
direction of $(-406, 352, -324)\kms$ \citep{Kogut93}.  The agreement
between the simulated and the observed velocity amplitude at $t=t_0$
is impressive given the crudeness of the initial conditions of the
simulation; however, the direction of the velocity vector deviates
somewhat from the observed orientation.

More recently, \citet{Tonry00} used the results from the Surface
Brightness Fluctuation Survey to model the local and large-scale
flows. Their best-fit model includes two attractors, one of which
having a best-fit location coincident with the Virgo cluster and the
other having a fit location slightly beyond the Centaurus cluster
(which is commonly referred to as the Great Attractor). Assuming an
attenuated power-law mass distributions for the two attractors, their
best-fit model has enclosed mass (in excess of background density) of
$7\times 10^{14}\Msun$ for Virgo and $9\times 10^{15}\Msun$ for the
Great Attractor within spheres centered on the attractors with radii
reaching the Local Group.  For comparison, the corresponding masses in
our simulation are $7\times 10^{14}\hinv\Msun$ for the Virgo and
$8\times 10^{15}\hinv\Msun$ for the Centaurus, in good agreement with
the observational estimate by Tonry et al.

We follow the trajectories of the above 11 particles up to the time 
$t=t_0+6t_H$.  The right panel of \Fig{virgo.eps} indicates that the
particles have traversed a comoving distance of $\sim 5\himpc$ towards the
direction of the Virgo cluster but have not fallen into it.  The physical
distance to Virgo (which is proportional to $a(t)$) would increase 
exponentially at late times. Neighboring particles show a similar 
behavior, so this conclusion is not likely to be affected by the random 
sampling of particles from the original data set described in 
\S~\ref{section:simulation}.  
By $t=t_0+6t_H$, the peculiar velocity field at moderate overdensities
is heavily damped (see \S~\ref{section:evolution}).
Most of the displacement towards Virgo takes place by $t=t_0+2t_H$. 
Note that the Virgo
cluster itself has also moved by a few comoving $\himpc$ in response to the
pull of the Great Attractor. We conclude that the Local Group will not be
gravitationally bound to the Virgo cluster,
given the initial conditions of the simulation that are matched to the 
observed galaxy distribution.

\subsection{The Milky Way and the Andromeda Galaxies}

The Local Group is a centrally concentrated system in which half of all
members lie within $\sim 450\kpc$ of its center-of-mass.  Most of the mass
is concentrated in the Andromeda and the Milky Way subgroups.
Unfortunately, the mass estimate of each subgroup suffers from large
uncertainties. For example, the mass of the Andromeda subgroup ranges from
a value of $M_{\rm A}=(13.3\pm1.8)\times 10^{11}\Msun$ \citep{Courteau99}
to a lower estimate of $7.0^{+10.5}_{-3.5}\times 10^{11}\Msun$
\citep{Evans00}. The mass of the Milky Way ranges from $M_{\rm
MW}=(8.6\pm4.0)\times 10^{11}\Msun$ \citep{Zaritsky99} to a larger value of
$19^{+36}_{-17}\times 10^{11}\Msun$ \citep{Wilkinson99}.  All of the
existing estimates are consistent with each other to within the quoted
errors.  Under most estimates, the total mass of the Local Group is close
to $M_{\rm LG}=(23\pm 6)\times 10^{11}\Msun$ \citep{Bergh99}.

We may evaluate the present-day overdensity of the Local Group,
$\delta_{\rm LG}$, using its total mass and the distance of $\ell=740\pm
40\kpc$ \citep{Binney98} between the Milky Way and Andromeda, yielding
$1+\del_{\rm LG} = 3M_{\rm LG}/[4\pi (\ell/2)^3 \bar{\rho}] =130h_0^{-2}$.
The present overdensity of the Local Group clearly exceeds the critical
collapse threshold of $\del_c=17.6$.

Adopting the measured radial component of the relative velocity between the
Andromeda and the Milky Way galaxies $v_r=-120\kms$ \citep{Binney87} and
ignoring its tangential component, it can easily be shown that the total
energy of the Local Group, $\frac{1}{2}\mu_{\rm LG} v_r^2 - GM_{\rm
A}M_{\rm MW}/\ell$, is negative, where $\mu_{\rm LG} = M_{\rm A}M_{\rm
MW}/(M_{\rm A}+M_{\rm MW})$ is the reduced mass of the Andromeda and the
Milky Way subgroups.  This result is obtained for either of the latter or the
former pair of mass values quoted above for $M_{\rm A}$ and $M_{\rm MW}$.  
The transverse component of the relative velocity is commonly assumed to be
much smaller than the radial velocity \citep{Einasto82}.  Based on the
radial separation and the relative velocity between the two galaxies, one
finds that they are likely to merge within a Hubble time.  A pedagogical
discussion on the dynamical evolution of the Local Group based on the least
action principle can be found in \citet{Peebles93}.  Dynamical evaporation
of stars out of the merger product can be neglected on the time-scale of
interest here \citep{Binney87}.

%%%%%%%%%%%%%%%%%%%%%%%%%%%%%%%%%%%%%%%%%%%%%%%%%%%%%%%%%%%%%%%%%%%%%%%

\section{Conclusions \& Discussions}
\label{section:conclusion}

We have simulated the future evolution of our cosmic neighborhood in a
universe dominated by a cosmological constant using an N-body code and
the initial conditions that were reconstructed from the observed galaxy
distribution of the {\it IRAS} 1.2 Jy survey.  We find that the
large-scale structure and the mass distribution of bound objects will
freeze in $\sim 30$ billion years from today due to the accelerated
expansion of the Universe.  The Local Group of galaxies will get
somewhat closer to the Virgo cluster of galaxies in comoving
coordinates, but will rapidly recede from Virgo in physical
coordinates.  However, the overdensity inside the Local Group is well
above the required threshold ($\delta_c=17.6$) for it to resist the
repulsive gravitational force of the cosmological constant. Therefore
the Milky Way and the Andromeda galaxies are likely to merge within a
Hubble time.

If the Universe is dominated by a cosmological constant, then the
Local Group is detached from the rest of the Universe and the physical
distance from us to all other systems that are not bound to the Local
Group will increase exponentially with time in the distant future.
Combining our simulations with the results of Loeb (2002), we predict
that when the age of the Universe will be $\sim 100$ billion years,
there will only be one massive galaxy for us to observe, namely the
merger product of the Andromeda and the Milky Way galaxies.  All other
systems that are not gravitationally bound to the Local Group will
exit through our event horizon and their images will fade rapidly
while remaining frozen on the time of their exit.

The precision of our quantitative results is limited by the accuracy
of the initial conditions of the simulation at $z=0$ compared to the
true mass density field in the nearby universe.  The locations of the
simulated clusters of galaxies are shifted by a few megaparsec
relative to the recent observational estimates, and in principle the
match between the two can be improved in the future by refining the
reconstruction method of the initial conditions of the simulation from
the observational data.  While we believe that the general predictions
derived in this paper are independent of the limited mass and spatial
resolution of our simulations, it would be appropriate to perform
in the future higher-resolution simulations with improved initial
conditions that match the observed galaxy distribution. Since
the observational uncertainties in the distance measurements are still
substantial, the precision of the observed galaxy distribution will
also get better with future observations.  The combination of both of
the above improvements will allow us to make more accurate predictions
for the future evolution of the nearby large-scale structure.

%%%%%%%%%%%%%%%%%%%%%%%%%%%%%%%%%%%%%%%%%%%%%%%%%%%%%%%%%%%%%%%%%%%%%

%\acknowledgments 
\vspace{1cm}

{\bf Acknowledgments}\\
We are grateful to the authors of Mathis et al. (2002) paper and the GIF
(German-Israel Foundation) collaboration for allowing us to use
their $z=0$ simulation output as the initial condition of our simulation.
In particular, we thank Volker Springel for providing us with the data in a
convenient format for the GADGET code, as well as for clearing out some of
our technical problems and questions at the initial stage of this work.  
This work was supported in part by the grants 5-7768 from NASA and 
AST-9900877, AST-0071019 from NSF for AL.

%%%%%%%%%%%%%%%%%%%%%%%%%%%%%%%%%%%%%%%%%%%%%%%%%%%%%%%%%%%%%%%%%%%%%%

%%%%%%%%%%%%%%%%%%%%%%%%%%%%%%%%%%%%%%%%%%%%%%%%%%%%%%%%%%%%%%%%%%%%%%

\end{document}